\documentclass[twoside,12pt]{article}
\usepackage{epsfig}

\def\Journal#1#2#3#4{{#1} {#2} (#4) #3 }
\def\AP{{\em Ann. Phys.}}
\def\AA{{\em Astron. \& Astrophys.}}

\def\AFZ{{\em Astrofizika}}

\def\APJ{{\em Astrophys. J.}}

\def\MNRAS{{\em MNRAS}}
\def\N{{\em Nature}}
\def\NPA{{\em Nucl. Phys.} A}
\def\NPB{{\em Nucl. Phys.} B}

\def\PLB{{\em Phys. Lett.} B}

\def\PRL{\em Phys. Rev. Lett.}

\def\PREP{\em Phys. Rep.}

\def\PRD{{\em Phys. Rev.} D}

\def\arx{{\em arXiv:}}

\newcommand{\be}{\begin{equation}}
\newcommand{\ee}{\end{equation}}
\newcommand{\bea}{\begin{eqnarray}}
\newcommand{\eea}{\end{eqnarray}}

\newcommand{\lsim}{\stackrel{\scriptstyle <}{\phantom{}_{\sim}}}
\newcommand{\gsim}{\stackrel{\scriptstyle >}{\phantom{}_{\sim}}}
\begin{document}

\title{ \vspace{1cm}
Asymmetric neutrino propagation in newly born magnetized strange stars: GRB
and kicks}
\author{J.\ Berdermann,$^{1}$ D.\ Blaschke,$^{2,3}$
H.\ Grigorian,$^{1,4}$ and D. N.\ Voskresensky$^{2,5}$ \\
\\
$^1$Institut f\"ur Physik der Universit\"at Rostock, D-18051 Rostock, Germany\\
$^2$Gesellschaft f\"ur Schwerionenforschung (GSI) mbH, D-64291 Darmstadt,
Germany\\
$^3$Bogoliubov Laboratory for Theoretical Physics at JINR, 141980 Dubna,
Russia \\
$^4$Department of Physics, Yerevan State University, 375049 Yerevan, Armenia\\
$^5$Moscow Institute for Physics and Engineering, 115409 Moscow,
Russia\\
}
\date{ }
\maketitle
\begin{abstract}
We consider the early cooling evolution of strongly magnetized strange stars
in a CFL phase with high gap $\Delta \gsim 100$~MeV.  
We demonstrate how this model  may explain main features of the gamma-ray burst
phenomena and also yield a strong star kick.  
The mechanism is  based on beaming of neutrino emission along the magnetic 
vortex lines. 
We show that for sufficiently high initial temperatures 
$T_0\sim 30$ to $60~$MeV and surface magnetic fields $B_s \sim 10^{15}$ to  
$10^{17}~$G, the energy release within the narrow beam is up to  $10^{52}$~erg 
with a magnetic field dependent time scale between $10^{-2}$~s (for a smaller 
magnetic field) to $10$~s.
The above mechanism together with the parity violation of the 
neutrino-producing weak interaction  processes in a magnetic field allow for 
the strange star  kick.  
The higher the magnetic field the larger is the stars kick velocity.
These velocities may cover the same range as observed pulsar kick velocities. 
\end{abstract}

\section{Introduction}
Compact stellar objects are subdivided into three groups:
usual neutron stars (having only hadronic
matter interiors and an extended crust), hybrid stars (with quark core,
hadronic shell and a crust similar to that in neutron stars) and strange stars 
(hypothetical self-bound objects composed of $u$, $d$ and $s$ quark matter 
with only a thin crust if any).
The possibility of strange stars has been extensively discussed in the 
literature, see \cite{Bombaci:2001uk} and Refs. therein.

Gamma ray bursts (GRB) are among  the most intriguing phenomena 
in the Universe, see \cite{Piran} and Refs. therein. 
If the energy is emitted isotropically, an energy release  of the order of  
$10^{53}$ to $10^{54}$~erg is needed to power the GRB. 
However, there is now compelling evidence that the gamma ray emission is not 
isotropic, but displays a jet-like geometry. 
In the case of beaming a  smaller energy, of the order of $10^{51}$~erg, would 
be sufficient to power the GRB \cite{Frayl}.
Although typically one may speak about a short GRB
with a time scale $t\sim 10^{-3}$ to $ 10^{-1}~$s and a long GRB 
with a time scale $t\sim ~1$ to $ 100$~ s,
energies, durations, time scales involved in a burst, pulse shape structures, 
sub-burst numbers, etc, vary so much that it is hard to specify a typical GRB. 

Many scenarios for the GRB have been proposed:
the ``collapsar'', or ``hypernova'' model
linking the GRB with ultra-bright type Ibc supernovae (hypernovae) 
and the subsequent black hole formation \cite{BB};
neutron star mergers \cite{ros}, or accretion of matter onto a black hole; 
strange star collisions \cite{Haensel:1991um};
the model  \cite{Usov} assuming a large surface magnetic field 
up to $ 10^{16}$~G of a millisecond pulsar 
produced in the collapse of a  $\sim 10^9~$G
white dwarf,
with a powerful pulsar wind, as the source 
of the GRB; the model
\cite{SW1} suggesting a $ee^+$ plasma wind between heated neutron stars
in close binary systems as consequence of the $\nu\bar{\nu}$ annihilation; 
the model \cite{RK} of a steadily accreting $\sim 10^6~$G 
white dwarf collapse to a millisecond pulsar
with a $\sim 10^{17}$~G interior toroidal field,
causing the GRB; an (isotropic) first order phase transition \cite{Bombaci} 
of a pure hadronic compact star to a strange star (see also \cite{Aguilera}); 
 asymmetric core combustion \cite{Lug}
in neutron stars due to the influence of the 
magnetic field generating an acceleration of the flame in the polar direction,
etc.

Pulsars are rotating neutron stars with rather high magnetic fields causing 
observable radio dipole signals.
Most of the known pulsars are born in the neighborhood of the galactic plane
and move away from it with natal kick velocities which are typically higher
than those of their progenitors \cite{Lyne94}.
This implies that the birth process of pulsars also produces their high
velocities and thus cannot be entirely isotropic \cite{Spruit:1998sg}.  
Up to now, the mechanisms driving this asymmetry are far from clear.
There are several hypotheses, ranging from
asymmetric supernova explosions \cite{lai01}, neutron star instabilities
\cite{colpi03,imshennik04} and magneto-rotational effects
\cite{moiseenko03,ardeljan04} to the model of an electromagnetic or neutrino
rocket \cite{lai01}.
It is also not clarified whether the distribution of pulsar kick
velocities is bimodal with a low-velocity component of $v \leq 100 ~{\rm km/s}$
(20 \% of the known objects) and a high-velocity component of
$v \geq 500 ~{\rm km/s}$ (80\%) as suggested by \cite{Arzoumanian:2002} or
whether it can be explained by a one-component distribution
\cite{Hobbs:2005yx}.
Most of the models are capable of explaining kick velocities of
$v \sim 100 ~{\rm km/s}$, but it is a nontrivial problem to explain
the highest measured pulsar velocities around $1600 ~{\rm km/s}$.

In this work we assume  that a strongly magnetized strange star has been
formed in the color superconducting color-flavor-locked (CFL) state with a 
large gap $\Delta (T)$, $\Delta (T=0)\gsim 100$~MeV, cf. 
\cite{Rajagopal:2000wf},
\footnote{Our results are also relevant for the mCFL phase \cite{IMTH}. 
We only need all quarks to be gapped with large gaps} as the result of a phase 
transition. 
The latter could be caused by the accretion of the matter from a companion 
star, the neutron star angular momentum decrease owing to the  gravitational 
and electromagnetic radiation after the supernova event had occurred, or during
the collapse of a magnetized white dwarf to the neutron star state, the 
proto-neutron star collapse to the new stable state, or by some other reason. 
We also can deal with a hybrid star instead of the strange star. However, cf. 
\cite{RK}, there is experimental evidence that the GRB  carries only a tiny 
baryon load of mass $\lsim 10^{-4}M_{\odot}$.
Therefore the  hadronic shell and the crust should be rather thin or there 
should be a special reason for a low baryon loading. 
We suggest to explain the beaming by the presence of a strong magnetic field.
{\em Thus the beaming and the low baryon loading stimulate us to 
conjecture about a magnetized strange star,} 
which has a tiny  hadron shell, if any, and only a thin crust.
Varying  the value of the surface magnetic field $B_s$, we search for an 
optimal configuration to explain the GRB characteristics  and to estimate a
range of velocities of the star kicks.
We use units $\hbar =c=1$.

\section{Vortices in  magnetic field, neutrino beaming and GRB}

{\bf{Vortex structures in the CFL phase of the strange star  in  the 
magnetic field.}}
Let us further assume that the bulk of the star is in the CFL phase
with no hadronic shell and a tiny crust generating a surface magnetic
field which may achieve the values
$B_s ~ \sim 10^{12} - 10^{16}~ {\rm G}$. 
Then, one may expect by flux conservation that
the inner magnetic field (at the crust-core interface)
$B_{in,s}= B_s (n_{in}/n_s)^{2/3}$
can reach the values $\gsim 10^{14} - 10^{18}~{\rm G}$, 
where $n_s$ 
and $n_{in}$
are the corresponding densities at the surface and in the core, respectively. 
Below we will vary  $B_{in,s}$
within the interval $10^{15}\lsim B_{in,s}\lsim 10^{17}$~G.
From the BCS relation 
the critical temperature of
the superconductivity is $T_c = 0.57~\Delta (T=0)$ and the temperature
dependent gap can be parameterized as \cite{LO02}
$\Delta (T)\simeq \Delta (T=0)[1-(T/T_c)^{3.4}]^{0.53}$. 
For initial temperatures
$T_0\simeq 30 - 60~$MeV and the gap $\Delta (T=0)\gsim 100$~MeV we have 
$T_0 < T_c$. 
Thus the strange star is formed in the superconducting phase.
Refs.
\cite{Blaschke:1999fy,Blaschke:1999qx,Sedrakian:2000kw,Giannakis:2003am}
have shown, that typical values of the Ginzburg-Landau parameter for
color superconductors are $\kappa_{GL}={\lambda}/{\xi}
\stackrel{>}{\sim} 1$. The coherence length $\xi_{{\rm CFL}}$ and
penetration depth $\lambda_{{\rm CFL}}$ for CFL matter can be
estimated in the weak coupling limit as 
\begin{eqnarray} \label{xila}
\xi_{{\rm CFL}} \simeq 0.3\biggl(\frac{100~{\rm MeV}}{T_c}\biggl)
\biggl( 1-\biggl(\frac{T}{T_c}\biggl)^{3.4}\biggl)^{-0.53}~{\rm fm}, ~~
\lambda_{{\rm CFL}} \simeq 2\biggl(\frac{3\sqrt{2}}{\sqrt{3g^2}}\biggl)
\biggl(\frac{300~{\rm MeV}}{\mu_q}\biggl)
\biggl(1-\biggl(\frac{T}{T_c}\biggl)^{3.4}\biggl)^{-0.53}~{\rm fm}~,
\end{eqnarray}
where we used the above $\Delta (T)$ dependence. 
In our case the strong coupling constant $\alpha_s =g^2/4\pi \sim 1$, and  the
quark chemical potential is $\mu_q \geq 350~$MeV.
The critical Ginzburg-Landau parameter is $\kappa_{GL}^c \simeq 
1/\sqrt{2}$ distinguishes between type I ($\kappa_{GL} < \kappa_{GL}^c$)
and type II ($\kappa_{GL} > \kappa_{GL}^c$) superconductors. 
Then for $T_c >T_c^{\rm I-II} \sim 15\div 20~{\rm MeV}$
we deal with a type II superconductor. 
The latter entails the existence of a mixed phase (normal vortices embedded in 
superconducting matter)  for $T<T_c$ in a broad interval of magnetic fields
$B_{c1}<B<B_{c2}$. 
The value $B_{c2}$ is $\simeq B_c \kappa_{GL}$, with $B_c \gsim 10^{18}$~G 
for $\Delta \gsim 100$~MeV.
Actually, even if $B<B_{c1}=B_c /\kappa_{GL}$, the magnetic field, in spite of 
the Meissner effect, cannot be expelled from the star interior within a 
pulsar lifetime \cite{BPPR}, thus being concentrated in vortices aligned 
along the magnetic axis (feasibly being parallel to the rotation axis). 
Therefore, the mixed state exists for all $B<B_{c2}$.
Making below only rough estimates we will assume the simplest  Abrikosov vortex
structure of the mixed phase.

Due to magnetic flux conservation the  number of vortices in the strange star 
interior is
$N_{vo} = \pi B_{in,s} R^2/\Phi_q$,
where $\Phi_q =6\Phi_0$ with the magnetic flux quantum 
$\Phi_0 =2\cdot 10^{-7}~$G$~ \mbox{cm}^2$ and the star radius $R\simeq 10$~km
\cite{Blaschke:1999fy,Blaschke:1999qx,Sedrakian:2000kw}.
Superconductivity is expelled from the vortex interior, $r\lsim \xi$,
where the matter is in the state of a strongly magnetized quark-gluon plasma.
The magnetic field decreases on the scale of the penetration
depth $\lambda_{\rm CFL}$ as $B\sim B_0~ \mbox{exp}(-r/\lambda_{\rm CFL} )$, 
for $r>\lambda_{\rm CFL}$,
cf. \cite{Lifshiz:1980,Blaschke:1999fy,Blaschke:1999qx,Sedrakian:2000kw}. 
Typical values of the magnetic field in the vortex center are 
$ B_0\sim \Phi_{q}/(\pi\lambda_{\rm CFL}^2 ) \sim  10^{18} - 10^{19}~$G, 
depending on $\lambda_{\rm CFL}$. 

In hadronic matter and in  metals, the pairing gap is rather 
small and the coherence length $\xi\propto 1/\Delta (T)$ is 
larger than the Debye screening length.
Therefore, charge neutrality is easily fulfilled for $r\lsim \xi$.
In the quark matter case we have opposite situation 
\cite{Voskresensky:2001jq,Voskresensky:2002hu}. 
For both, quarks and electrons, $\lambda_D =\nu\lambda_{CFL}$ with typical
values of $\nu \sim 3 - 5$. 
The presence of the Coulomb field diminishes the difference between the 
chemical potentials of quarks of different species since one does not need to 
fulfill the charge neutrality condition locally at distances $r\ll \lambda_D$
but rather globally at $r \gsim  \lambda_{D}$. 
This implies that the cylindrical volume within distances $r<\lambda_D$
from the vortex center is possibly filled by  a 2SC+s phase plus a normal 
phase \cite{Blaschke:2005uj}, also 
characterized by a still rather high magnetic field $B\sim 10^{17}$~G.  
In case of the two-flavor superconducting (2SC) phase quarks of two colors, 
e.g., green and red, are paired with a large gap, whereas blue quarks are 
unpaired or paired with only a small gap $\lsim 1$~MeV \cite{Rajagopal:2000wf}.
The 2SC+s phase emits neutrinos with a still high rate 
of about 1/3 of the emissivity of normal quark matter. 
In this phase at
densities under consideration typical values of the electron fraction $Y_e
=n_e /n_b$, where $n_e$ and $n_b$ are electron and baryon densities, are
$Y_e^{vo} \sim 10^{-2}$ (for $r\sim \lambda_D$). 
The electron fraction in the CFL phase at $T=0$ vanishes
\cite{Rajagopal:2000wf}, but at finite temperatures $Y_e^{CFL}(T)\neq 0$ 
due to thermal $ee^+$ excitations.

The typical distance between vortices is $d = ( \Phi_q / B_{in,s})^{1/2}.$
The region near the magnetic vortex, where the
magnetic field decreases by up to two orders of magnitude
compared to the central value $B_0$, has the volume
$V_{vo} \sim 2\pi~R(\nu \lambda_{CFL})^2 \sim 10^{21}~ {\rm fm^3}.$
The fraction of the total vortex volume $N_{vo}V_{vo}/V$
($N_{vo}$ is the total number of vortices, $V$ is  the star volume) varies
from several $\cdot 10^{-7}$ to several $\cdot 10^{-3}$ for $B_{in,s}$ from 
$\sim 10^{13}$~G to $10^{17}$~G, respectively. 
Therefore the condition $d\gg \lambda_D$ is safely fulfilled.

{\bf{Goldstone transport}}.
The CFL (and mCFL) phase is also characterized by the presence of (almost) 
massless weakly interacting Goldstone excitations (like phonons).
For $T\gg T_{opac}^{G}\simeq 2 - 3 ~$ MeV of our interest 
the mean free path of Goldstones is much shorter than  the star radius 
$\sim 10~$ km, cf. \cite{Shovkovy:2003bf}, and the typical time of their 
transport to the surface is very large.
In the presence of a strong magnetic field generating  vortices Goldstones
cannot pass from CFL matter to the 2SC+s matter and to the
normal quark matter phase of the vortex core, where they cannot exist. 
Thus, the mean free path of Goldstone excitations $\lambda_{GB}^{vo}$ in 
presence of vortices is still much shorter than in a homogeneous CFL strange 
star  matter.
For magnetic fields of our interest one finds 
$ \lambda_{GB}^{vo} \sim d^2 /\xi \approx 6 \times (10^2 - 10^6)$ fm. 
Thus Goldstones may efficiently equilibrate the temperature between the
vortex interior and exterior regions. 

{\bf{Neutrino radiation and beaming}}. 
The emissivity of  the quark direct Urca (QDU) process in normal quark matter 
in absence of the magnetic field is very high  \cite{Iwamoto:1980eb}
\begin{equation} \label{QDU}
\epsilon^{QDU}_{\nu} \simeq 2 \times 10^{26}~\alpha_s~u~Y_e^{1/3}~T^6_9~ 
{\rm erg~cm^{-3}~s^{-1}}.
\end{equation} 
Correspondingly, the neutrino mean free path  in normal quark matter 
becomes very short
\begin{equation} \label{MFP}
\lambda_{\nu}^{QDU} = 
3\times 10^6~\alpha_s^{-1}~u^{-1}~Y_e^{-1/3}~T_9^{-2}~{\rm cm}. 
\end{equation}
For $T\ll p_{Fe}$ one $Y_e =p_{Fe}/p_{Fb}$, where 
$p_{Fi} =(3\pi^2 n_i)^{1/3},~i=e,p$
and $u = n_b/n_0$ is the compression. 

In superfluid matter the suppression factor of the emissivity is roughly 
\cite{Blaschke:2000dy1}
\begin{equation} \label{supp}
\zeta_{DU}\simeq  ~
~\exp(-\Delta (T)/T), \quad T<T_c,
\end{equation}
and the
enhancement factor of the mean free path is then $\zeta_{DU}^{-1}$.
In the CFL phase $p_{Fe} =0$ but there exist thermal $ee^+$ excitations. 
Then  $p_{Fe}$ should be replaced by $T$ and we may use
above expressions (including approrpiate suppression and enhancement factors) 
with $Y_e^{CFL}\simeq T^3/(3\pi^2)n_b$.
The emissivity remains virtually unaltered in the presence of a magnetic field 
when many Landau levels are occupied
\cite{Bandyopadhyay:1998aq}
and it is increased by $10 - 100$ times if only the first Landau level 
is filled. 
However, the latter  occurs only for  $B\gsim 10^{18}$~G. 
With the field values which we have assumed this condition is
satisfied  only in the region of the
vortex cores, $r\lsim \xi$, of a tiny volume. 
In the remaining part of the vortex  volume $V_{vo}$ ($\xi <r<\nu\lambda_D$,
$\xi \ll \nu\lambda_D$)
one can roughly neglect the influence of the magnetic field on eq. (\ref{QDU}).
Also multiplying (\ref{QDU}) by a factor $1/3\leq s<1$ 
we may estimate the emissivity of the 2SC+s +normal phase since due to the 
pairing in the 2SC phase the contribution of the red and green
quarks is exponentially suppressed according (\ref{supp}).
Also in the region between the vortices occupied by CFL phase with the volume 
$V-N_{vo}V_{vo}$, the emissivity of the direct Urca process is suppressed by
(\ref{supp}).
The neutrino mean free path increased by this factor exceeds the
radius of the star for temperatures below  $T_{opac}^{CFL}\sim 20$ MeV (for
$\Delta (T=0)\sim 150$~MeV),
whereas for the normal quark matter one gets $T_{opac}^{norm}\lsim  $ few
MeV, cf. \cite{Haensel:1991um}.
For higher gaps the value $T_{opac}^{CFL}$ still increases.

For $T>T_{opac}^{CFL}$ the heat transport is governed by the transport 
equation, $C_V \dot{T}=\kappa \Delta T$, $C_V$ is the total specific heat of 
the matter and
$\kappa = \kappa_{\nu}+\kappa_{G}$ is the total heat conductivity.
Since Goldstones (phonons) are efficiently captured by vortices one has 
$\kappa_{\nu}\gg \kappa_{G}$, and thus $\kappa \simeq \kappa_{\nu}$. 
A typical time scale for the energy transport to the surface 
is $t_{tr}\sim R^2 C_V /\kappa $. 
Since $\kappa_{\nu}\sim C_V \lambda_{\nu}$,
the energy transport time 
 is $t_{tr}\sim R^2 /(\lambda_{\nu})$. 
Using the above value $\lambda_{\nu}^{QDU}$
for $T\sim 20 - 40~$MeV, we get $t_{tr}\sim 10^{-4} - 10^{-1}~$s
(estimate is done for
$\Delta (T=0)\simeq 150$~MeV). 
For that time the star essentially losses its initial thermal  energy. 
If initially (for $T_0 =40$~MeV, $\Delta \gsim 100$~MeV) the thermal energy is
mainly concentrated in massless excitations, the energy falls by a factor 
$2^4$ for $T\simeq T_{opac}^{CFL}\simeq 20~$MeV. 
After the transport time $t_{tr}$ the CFL regions become transparent for 
neutrinos.

For $T< T_{opac}^{CFL}$, the prominent difference between the neutrino mean 
free path in the 2SC+s+normal matter on the one hand and the CFL quark matter 
on the other hand  leads to anisotropic neutrino emission via the
direct (no rescatterings) QDU reaction  from the
star within a cone of the temperature-dependent opening angle 
$\theta_{\nu}(T) \sim \tilde{\lambda}_{\nu}(T)/R$ around the magnetic axis, 
where $\tilde{\lambda}_{\nu}(T)=\lambda_{\nu}(T)~V /(N_{vo}~V_{vo})$. 
Thus after $t > t_{tr}$ the star begins to radiate the remaining energy 
$\sim 10^{52}$~erg within a narrow beam cone. 

{\bf{Cooling evolution. Numerical results}}. The cooling evolution  for
$t>t_{tr}$
within the opening angle
can be described by inverting the solution of
\begin{equation}\label{timecool} 
t = -\int\limits_{T_{opac}^{CFL}}^{T(t)} 
~{\rm d}T'~C_V(T')
/L(T'),
\end{equation}
where 
\begin{equation} \label{CVtot}
C_V(T) = \{ [1-{N_{vo}V_{vo}}/{V}]~\zeta_{DU}+
s {N_{vo}V_{vo}}/{V}\} C_V^q(T)+C_V^{ex}(T), 
\end{equation}
$C_V^q(T)\simeq 10^{39} u^{2/3} (R/10\mbox{km})^3 T_9 \,\,\mbox{erg/K}$
is the specific heat of the normal quark matter
and $C_V^{ex}(T)\simeq
3\cdot 10^{32} g^{ex} (R/10\mbox{km})^3 T_9^3\,\,\mbox{erg/K}$ is the
contribution of (almost) massless Goldstone
excitations and electrons and positrons 
(for the effective value of the degeneracy factor $g^{ex}\simeq 11$, we have
assumed $v_G\simeq 1/\sqrt{3}$ for the velocity of Goldstones, the maximum 
Fermi velocity for relativistic fermi liquids. 
If $v_G$ were $\simeq 1$, we would get $g^{ex}\approx 3$.).
The luminosity  is
\begin{equation}\label{solL}
L (T)= [1-{\rm cos} ~\theta_{\nu}(T)] L_0(T) \simeq
L_0(T) \theta_{\nu}^2 (T) /2 
\end{equation}
that takes into account the effect of the neutrino beaming.
Here the isotropic luminosity is
\begin{eqnarray}\label{f7.9+}
L_0(T) = N_{vo}V_{vo}
s\epsilon_{\nu}^{QDU}(T, Y_e^{vo})
+ (V- N_{vo} V_{vo})  
\epsilon_{\nu}^{QDU}(T, Y_e^{CFL} (T) )~\zeta_{DU}.
\end{eqnarray}
We take the electron fraction $Y_e^{vo}\simeq 10^{-2}$ for the 2SC+s + normal
phase and  a thermally produced electron fraction $Y_e^{CFL}(T)$
for the CFL phase.

Here the Goldstones play an essential  role since they rapidly 
(in a microscopic time) equilibrate the temperature between the vortices 
(of volume $N_{vo}V_{vo}$) and their surrounding. 
In Eq. (\ref{timecool}) we neglected this very short time assuming  an 
instantaneous heat transport from the  hotter CFL regions of vortex exteriors 
to the regions of the vortex volume which  then cool down 
by the direct neutrino radiation  within the beaming angle. 

\begin{figure}[h]
\begin{center}
\begin{minipage}[t]{12cm}
\vspace{-10mm}
\epsfig{file=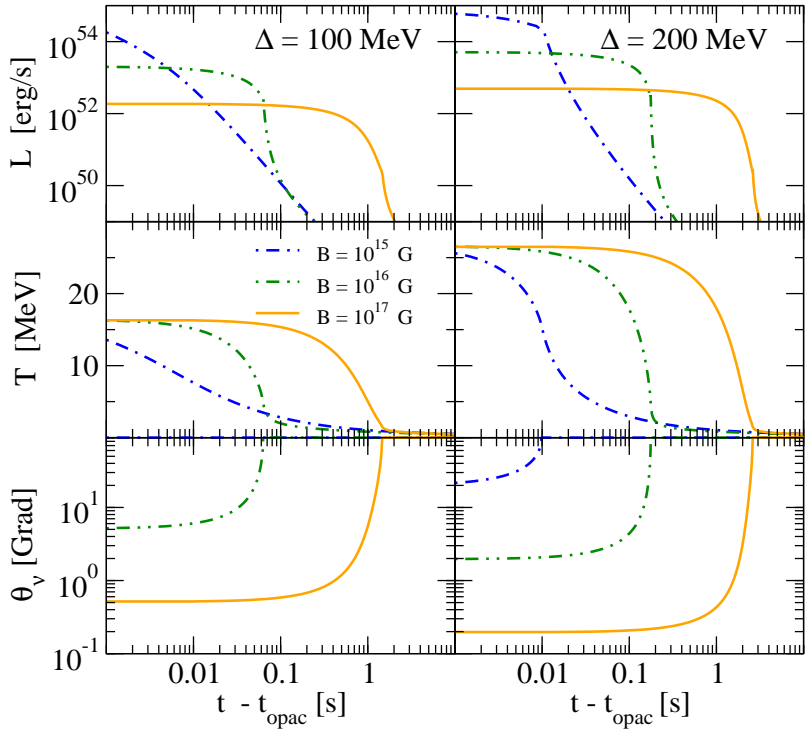,height=14cm,width=11.5cm,angle=-90}
\end{minipage}
\begin{minipage}[t]{6cm}
\caption{Time evolution of neutrino luminosity, temperature and neutrino
beaming angle for diquark gaps $\Delta=100$ MeV (left panel) and 
 $\Delta=200$ MeV (right panel), for different magnetic fields  $B=B_{in,s}$.
\label{fig1}}
\end{minipage}
\end{center}
\end{figure}

Results for the time evolution of neutrino luminosity, temperature and the 
neutrino beaming angle are shown in Fig. \ref{fig1}. 
They demonstrate the effect of the cooling 
delay due to neutrino beaming in a strong magnetic field.
For gaps $\sim 100 - 200 $ MeV and $B_{in,s}\sim 10^{15}$G we get 
$\theta_{\nu} \sim 10$ grad and the  
energy $\sim 10^{52}$~erg that remained in the star after the neutrino
transport era has passed is radiated in the beam during several 
$\sim 10^{-2} - 10^{-1} $~s.
Decreasing further the magnetic field one would get  increase  of the opening 
angle $\theta_{\nu}$ angle finally yielding 
inappropriate smearing of the beaming. 
For $B_{in,s}\sim 10^{17}$G we obtain  a narrow beam, $\theta_{\nu} \sim
0.1 - 1$ grad, and the initial star
energy is radiated in the beam during several seconds.

{\bf{Neutrino energy conversion to photons, GRB}}.
The neutrino/antineutrino collisions produce
$ee^+$ pairs outside the star,
which efficiently
convert to photons. 
We use the result of \cite{Haensel:1991um} for the 
$\nu\bar{\nu}\rightarrow ee^+$ conversion rate:
\begin{eqnarray}\label{Haenee}
\dot{E}_{ee^+}\sim 5\cdot 10^{32} \,T_{s,9}^9 (R/10 \mbox{km})^3 
\quad~{\rm erg~s^{-1}} 
\end{eqnarray}
to estimate that for a surface temperature
$T_{s}\sim 10 - 30$~ MeV 
most of the initial  thermal energy can be converted in
$ee^+$ for  an appropriate  short time step $t\sim 10^{2} - 10^{-2}~$s. 
The advantage of the strange star model is that the strange star has a thin 
hadron shell and a tiny crust, if any. 
In this case we may assume that the surface temperature is of the same order 
of magnitude as the internal temperature.
Note that  in the presence of beaming already 
$E_\gamma\sim \mbox{several}\cdot 10^{51}~$erg produced on a time scale
 $t\sim 10^{-3} - 10^3~$s
could be sufficient to explain GRB.
Thus, above estimates are very optimistic for the GRB model dealing with  a 
magnetized quark star, if $T_{s}\gsim 10~$MeV.

Moreover, $ee^+$ pairs are accelerated in the strong magnetic field of the
strange star exterior, increasing the hard component of the $X$ ray spectrum,
in coincidence with experimental findings.

\vspace{-1cm}

\section{Strange star kicks from parity violation}
\begin{figure}[h]
\begin{center}
\begin{minipage}[t]{12cm}
\vspace{-5mm}
\epsfig{file=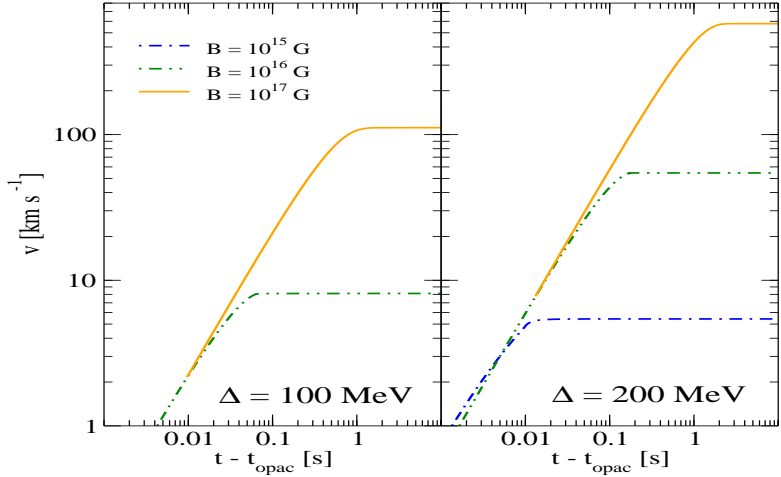,height=14.5cm,width=7.5cm,angle=-90}
\end{minipage}
\begin{minipage}[t]{6cm}
\caption{Time evolution  of the pulsar kick velocity for two values of 
the diquark gaps $\Delta=100$ MeV (left panel) and 
 $\Delta=200$ MeV (right panel), for different magnetic fields $B=B_{in,s}$.
\label{fig2}}
\end{minipage}
\end{center}
\end{figure}

{\bf{Range of kick velocities}}.
Parity non-conservation in the weak interaction neutrino induced direct 
processes in presence  of a magnetic field leads
to a violation of reflection symmetry since the neutrino flux is a polar vector
while the magnetic field is an axial one.
The magnitude of this asymmetry has been estimated considering different
reactions in neutron stars as the modified Urca reaction \cite{LP87}, see 
also \cite{M}; $\beta$ decays on the pion condensate \cite{P88};  
formation and breaking of nucleon pairs  \cite{P89}; coherent neutrino
electron scattering on nuclei \cite{Vilenkin:1995um}; neutrino-polarized
neutron elastic scattering; polarized electron
capture  \cite{Horowitz:1997mk};  the
$\nu\rightarrow \nu ee^+$ process \cite{KM97}; the reaction
$\nu\bar{\nu}\rightarrow  ee^+$; etc. 
Some of these reactions produce an asymmetry factor up to
$A_\nu\sim 10^{-4}~B_{14}$, where $B_{14}=B/10^{14}$ G. 
In our case of the strange star in the CFL phase with a strong magnetic field 
the direct QDU process and  formation and breaking of quark pairs are relevant 
as well as the reaction $\nu\bar{\nu}\rightarrow  ee^+$ near the star surface.
Here we assume that the QDU processes for $T < T_{opac}^{CFL}$ within the 
beaming angle (in accordance with the no go theorem
\cite{Vilenkin:1995um,Kusenko:1998yy})
produce an asymmetry factor of the same order of magnitude as has been 
estimated in the literature.
This $A_\nu$ factor then leads to a net momentum transfer from the neutrino
flux to the star of mass $M$ resulting in a time-dependent kick velocity
\begin{eqnarray}
v(t)&=&
\frac{A_\nu~}{M } \int_{t(T_{opac}^{CFL})}^t dt' L(T)~,
\end{eqnarray}
which saturates at the magnetic field dependent asymptotic value
\begin{eqnarray}
v\simeq 1.7 \cdot B_{14} (M/M_\odot)^{-1} ~{\rm km~s^{-1}}, \quad \mbox{for}
  \quad A_\nu \simeq 10^{-4}~B_{14},
\end{eqnarray} 
as soon as the beaming ceases. Numerical results are shown in Fig. \ref{fig2} 
for a typical strange star with $M=1.4~M_\odot$ and $R=10$ km. 
In order to obtain large star kicks with  $v\sim 10^3$ km s$^{-1}$, magnetic 
fields
$B_{in,s}\sim 10^{17}$ G  and large gaps $\Delta \sim 200$~MeV are required. 

\vspace{-1cm}

\section{Conclusions} 
\label{conclusion}

We have shown that in the presence of a strong magnetic field
$B_{in,s} \sim 10^{15} - 10^{17} $~G, the early cooling evolution of
strange  stars with a color superconducting quark matter core in the CFL
phase right after a short transport era is characterized by anisotropic 
neutrino emission, collimated within a beaming angle $\theta_{\nu}$ around 
the magnetic axis.
Initial temperatures of the order of $T_0 \sim 30 - 60~$MeV allow for the 
energy release $\sim 10^{53}$~erg. 
An energy  $\sim 10^{52}$~erg is released within a narrow beam on
time scales from $10^{-2}$~s to $10$~s, appropriate for the phenomenology 
of both short and long GRB.
Exploring the beaming mechanism one may obtain a wide range of strange star 
kick velocities up to $10^{3} ~{\rm km/s}$, in dependence on the magnetic 
field, pairing gap, radius, mass and burst duration of the star.
The range of strange star kick velocities is thereby in a qualitative 
agreement with recent observational data on pulsar kick velocity  distribution.
All our estimates are very rough and essentially vary with the parameters of 
the model. 
Thus we just have shown a principal possibility of the application of
the model to the GRB and kick description.
A more systematic investigation, together with a consistent modeling of the
compact star structure will be developed along the lines described in this
contribution \cite{Berdermann}.

\subsection*{Acknowledgements}
J.B. and D.B. acknowledge support by DAAD grant No. DE/04/27956
and by the DFG Graduate School 567
``Strongly Correlated Many-Particle Systems'' at University of Rostock.
The work of D.N.V. has been supported by DFG grant
436 RUS 113/558/0-3 and by an RF President grant NS-5898.2003.02.
H.G has been supported by DFG under grant 436 ARM 17/4/05.
D.B. is grateful for the invitation to participate at
the School for Nuclear Physics in Erice and for the support by DFG.


\begin{thebibliography}{99}
\itemsep -2pt
\bibitem{Bombaci:2001uk}
  I.~Bombaci,
 \Journal{\em  Lect.\ Notes Phys.\ }{578} {253}{2001}

\bibitem{Piran} T. Piran, \Journal{\PREP}{314}{575}{1999};
 \Journal{\PREP}{333}{529}{2000}

\bibitem{Frayl} D.A. Frayl et al., 
\Journal{\APJ}{562}{L55}{2001}

\bibitem{BB}
G.E. Brown, C.H. Lee, R.A.M.J. Wijers, H.K. Lee, G. Israelian, and
H.A. Bethe, 
\Journal{\em New Astronomy}{5}{191}{2000}
\bibitem{ros}
S. Rosswog, and E. Ramirez-Ruiz, 
\Journal{\em AIP Conf. Proc.}{727}{361}{2004}; astro-ph/0309201

\bibitem{Haensel:1991um}
  P.~Haensel, B.~Paczynski and P.~Amsterdamski,
\Journal{\APJ}{375}{209}{1991}

\bibitem{Usov} V. Usov, \Journal{\em ASP Conf. Series}{190}{153}{1999}

\bibitem{SW1} J.R. Salmonson, and J.R. Wilson, \Journal{\APJ}{578}{310}{2002} 
\bibitem{SW2} J.R. Salmonson, and J.R. Wilson, \Journal{\APJ}{517}{859}{1999}
\bibitem{RK}
M.A. Ruderman, L. Tao, and W. Kluzniak, \Journal{\APJ}{542}{243}{2000}  

\bibitem{Bombaci} I. Bombaci, \Journal{\NPA}{754}{335}{2005}

\bibitem{Aguilera}
D.~Aguilera et al.,
\Journal{\em NATO Science Series}{197}{377}{2005}

\bibitem{Lug}
G. Lugones, C.R. Ghezzi, E. de Gouveia Dal Pino, and J. Horvath,
\Journal{\APJ}{581}{L101}{2002}

\bibitem{Lyne94}
A.G.~Lyne, and D.R.~Lorimer,
\Journal{\N} {369}{127}{1994}

\bibitem{Spruit:1998sg}
H.~Spruit and E.~S. Phinney,
\Journal{\N}{393}{139}{1998}

\bibitem{Hobbs:2005yx}
G. Hobbs, D.~R. Lorimer, A.~G. Lyne, and M.~Kramer,
\Journal{\MNRAS}{360}{963}{2005}

\bibitem{lai01} D. Lai, D.F. Chernoff, and J.M. Cordes,
\Journal{\APJ} {549}{1111}{2001}

\bibitem{colpi03} M. Colpi and I. Wasserman,
\Journal{\APJ} {581}{1271}{2002}

\bibitem{imshennik04} V.S. Imshennik and O.G. Ryazhskaya,
\Journal{\em Astron. Lett.} {29}{831}{2004}

\bibitem{moiseenko03} S.G. Moiseenko, G.S. Bisnovatyi-Kogan and N.V. Ardeljan,
\Journal{\em Proc. of IAU Coll.}{192}{}{2003}

\bibitem{ardeljan04}  N.V. Ardeljan, G.S. Bisnovatyi-Kogan, K.V. Kosmachevskii,
and S.G. Moiseenko,
\Journal{\em Astrofizika} {47}{37}{2004}

\bibitem{Arzoumanian:2002}
Z.~Arzoumanian, D.F. Chernoff, and J.~M. Cordes,
\Journal{\APJ}{568}{289}{2002}

\bibitem{Rajagopal:2000wf}
K.~Rajagopal and F.~Wilczek, 
\Journal{\em At the Frontier of Particle Physics}{3}{2061}{2001}

\bibitem{IMTH}
K. Iida, T. Matsuura, M. Tachibana, and T. Hatsuda,
\Journal{\PRL}{93}{132001}{2004}

\bibitem{LO02}
L. Lindbom and B.J. Owen, \Journal{\PRD}{65}{063006}{2002}

\bibitem{Blaschke:1999fy}
  D.~Blaschke, D.~M.~Sedrakian and K.~M.~Shahabasian,
   \Journal{\AA}{350}{L47}{1999}

\bibitem{Blaschke:1999qx}
  D.~Blaschke, T.~Kl\"ahn and D.~N.~Voskresensky,
\Journal{\APJ}{533}{406}{2000}

\bibitem{Sedrakian:2000kw}
  D.~Sedrakian, D.~Blaschke, K.M.~Shahabasyan and D.N.~Voskresensky,
\Journal{\AFZ}{44}{443}{2001}

\bibitem{Giannakis:2003am}
I.~Giannakis and H.-C. Ren.
\Journal{\NPB}{669}{462}{2003}

\bibitem{BPPR}
G.~Baym, C.~Pethick, D.~Pines, and M.~Ruderman,
\Journal{\N}{224}{872}{1969}

\bibitem{Lifshiz:1980}
E.M. Lifshiz and L.P. Pitaevsky,
{\em Statistical Physics, IX}, Akademie Verlag Berlin, (1980)

\bibitem{Voskresensky:2001jq}
D.N. Voskresensky, M.~Yasuhira, and T.~Tatsumi,
\Journal{\PLB}{541}{93}{2002}

\bibitem{Voskresensky:2002hu}
D.N. Voskresensky, M.~Yasuhira, and T.~Tatsumi,
\Journal{\NPA}{723}{291}{2003}

\bibitem{Blaschke:2005uj}
  D.~Blaschke, S.~Fredriksson, H.~Grigorian, A.~\"Oztas and F.~Sandin,
\Journal{\PRD}{72}{065020}{2005}

\bibitem{Shovkovy:2003bf}
I.A. Shovkovy and P.J. Ellis,
\Journal{\arx}{hep-ph/0303073}{}{2003}

\bibitem{Iwamoto:1980eb}
  N.~Iwamoto,
  \Journal{\PRL}  {44}{1637}{1980};
\Journal{\AP}{141}{1}{1982}

\bibitem{Blaschke:2000dy1}
D.~Blaschke, H.~Grigorian, and D.N. Voskresensky,
\Journal{\AA}{368}{561}{2001}

\bibitem{Bandyopadhyay:1998aq}
D.~Bandyopadhyay, S.~Chakrabarty, P.~Dey and S.~Pal,
\Journal{\PRD}{58}{121301}{1998}

\bibitem{LP87}
Y.M. Loskutov and K.V. Parfenov,  
\Journal{\em Vestn. Mosk. Univ., Fiz.}{30}{3}{1989}

\bibitem{M} 
A.B. Migdal, E.E. Saperstein, M.A. Troitsky, and D.N. Voskresensky,
\Journal{\PREP}{192}{179}{1990}

\bibitem{P88}
 K.V. Parfenov,
\Journal{\em Sov. J. Nucl. Phys.}{48}{651}{1988}

\bibitem{P89}
K.V. Parfenov,
\Journal{\em Sov. J. Nucl. Phys.}{49}{1126}{1989} 

\bibitem{Vilenkin:1995um}
  A.~Vilenkin,
\Journal{\APJ}{451}{700}{1995}

\bibitem{Horowitz:1997mk}
C.~J. Horowitz and J.~Piekarewicz,
\Journal{\NPA}{640}{281}{1998}

\bibitem{KM97}
A.V. Kuznetsov and  N.V. Mikheev, \Journal{\PLB}{394}{123}{1997}

\bibitem{Kusenko:1998yy}
  A.~Kusenko, G.~Segre and A.~Vilenkin,
\Journal{\PLB}{437}{359}{1998}

\bibitem{Berdermann}
J. Berdermann, D. Blaschke, H. Grigorian and D.N. Voskresensky,
{\em in preparation} 

\end{thebibliography}
\end{document}